\documentclass[sigconf,nonacm]{acmart}

\AtBeginDocument{%
  }

\begin{document}

\title{IDCIA: Immunocytochemistry Dataset for Cellular Image Analysis}

\author{Abdurahman Ali Mohammed}
\authornote{Both authors contributed equally.}
\authornote{Abdurahman Ali Mohammed, Wallapak Tavanapong, and Azeez Idris are with the Department of Computer Science at Iowa State University, Ames, IA 50011 USA.}
\email{abdu@iastate.com}
\orcid{0000-0003-2538-4736}
\affiliation{%
  \institution{Iowa State University}
  \streetaddress{P.O. Box 50011}
  \country{USA}
  \postcode{50011}
}

\author{Catherine Fonder}
\authornotemark[1]
\authornote{Catherine Fonder and Donald S. Sakaguchi
are with the Department of Genetics, Development, and Cell Biology (GDCB),
Molecular, Cellular, and Developmental Biology Program (MCDB), and the
Neuroscience Program, Iowa State University, Ames,
IA 50011 USA.}
\authornote{
 Nanovaccine Institute, Iowa State University, Ames, IA 50011, USA.
}
\email{cfonder@iastate.edu}
\affiliation{%
  \institution{Iowa State University}
  \streetaddress{P.O. Box 50011}
  \country{USA}
  \postcode{50011}
}

\author{Donald S. Sakaguchi}
\authornotemark[3]
\authornotemark[4]
\email{dssakagu@iastate.edu}
\affiliation{%
  \institution{Iowa State University}
  \country{USA}}

\author{Wallapak Tavanapong}
\authornotemark[2]
\email{tavanapo@iastate.edu}
\affiliation{%
  \institution{Iowa State University}
  \country{USA}
}
\author{Surya K. Mallapragada}
\authornote {Surya K. Mallapragada is with the Department of Chemical and Biological Engineering at Iowa State University, Ames, IA 50011 USA.}
\authornotemark[4]
\email{suryakm@iastate.edu }
\affiliation{%
  \institution{Iowa State University}
  \country{USA}
}
\author{Azeez Idris}
\authornotemark[2]
\email{aidris@iastate.edu}
\affiliation{%
  \institution{Iowa State University}
  \country{USA}
}

\renewcommand{\shortauthors}{Mohammed et al.}

\begin{abstract}
 We present a new annotated microscopic cellular image dataset to improve the effectiveness of machine learning methods for cellular image analysis. Cell counting is an important step in cell analysis. Typically, domain experts manually count cells in a microscopic image. Automated cell counting can potentially eliminate this tedious, time-consuming process. However, a good, labeled dataset is required for training an accurate machine learning model. Our dataset includes microscopic images of cells, and for each image, the cell count and the location of individual cells. The data were collected as part of an ongoing study investigating the potential of electrical stimulation to modulate stem cell differentiation and possible applications for neural repair. Compared to existing publicly available datasets, our dataset has more images of cells stained with more variety of antibodies (protein components of immune responses against invaders) typically used for cell analysis. The experimental results on this dataset indicate that none of the five existing models under this study are able to achieve sufficiently accurate count to replace the manual methods. The dataset is available at https://figshare.com/articles/dataset/Dataset/21970604.
\end{abstract}



\maketitle

\section{Introduction}
Cell biology is a sub-discipline of biology where the structure and physiological functioning, and interaction of cells are studied \cite{bradshaw_cell_2016}. Cells are examined under a microscope and imaged at a high resolution. In immunocytochemistry (ICC), different antibodies are used to visualize the presence of particular proteins to identify specific cell types in a given sample. Cell analysis involves a wide range of tasks, such as counting cells and measuring and evaluating cell state (e.g., shape, motility), cell health, and cell growth. Cell biology is closely intertwined with other fields, such as neuroscience, genetics, and molecular biology. One fascinating application area of cell biology is research for the potential diagnosis and treatment of diseases. The research in this area is full of potential and possibilities that could improve quality of life.

Deep Neural Networks (DNNs) have been applied in the analysis of microscopic cell images, including cell counting \cite{paul_cohen_count-ception_2017,xie_microscopy_2018}, segmentation \cite{al-kofahi_deep_2018,ghaznavi_cell_2022,hiramatsu_cell_2018,morelli_automating_2021}, and detection \cite{WANG2022102270,jiang,Fujita_2020_ACCV}. Given an input image, cell counting provides the number of cells in the image. In contrast, cell segmentation finds the contours of individual cells, separating them from each other and the background. On the other hand, cell detection localizes a cell by drawing the smallest rectangle around each cell in the input image. The advantages of DNNs over traditional machine learning methods are that DNNs automatically extract important properties (features) of the object of interest and use them to perform the intended task. However, the major drawback of DNNs is that it requires a large high-quality labeled dataset for accurate predictions. Existing DNN methods for cell counting can be broadly categorized into two groups: detection-based and regression-based categories.

The detection-based category undertakes the counting task by first detecting individual cells (contours, bounding boxes, or centroids of the cells) in a given image and counting the detected cells to obtain the final cell count \cite{morelli_automating_2021,khan_deep_2016}. These methods hinge on the availability of the annotated ground truth of the bounding box or a centroid of a cell. The methods are also dependent on the characteristics of the microscopic input images. In particular, detection-based methods fail to offer good performance when there is a high occlusion in the images. The regression-based category \cite{paul_cohen_count-ception_2017,xie_microscopy_2018} predicts the cell count without detecting individual cells. Some of these methods use only the ground truth cell count for each training image for training. Other methods predict a corresponding density map for a given image and obtain the final count from the predicted density map.

Our team examines cellular images taken after electrical stimulation experiments on stem cells for cell differentiation. Cell differentiation is the process in which an unspecialized cell develops and matures to become a specialized cell. Electrical stimulation of stem cells is potentially useful for stem cell therapy in patients with nerve injuries. Cell counting is an important step toward determining an appropriate amount of electrical voltage and stimulation duration to be applied. To perform the electrical stimulation, cells are placed on the surface of a scaffold, which are structures providing support for cells to grow within an interdigitated electrode region. Then the voltage is applied to the electrode pads of the scaffold, which are structures providing support for cells to grow. During an electrical stimulation experiment, cells exhibit changes in size, shape, and energy requirement \cite{das_electrical_2017,uz_determination_2020,uz_development_2019}. Following electrical stimulation, immunocytochemistry (ICC) is performed to measure the effect of the stimulation on the cells. Different antibodies are used during the ICC process to identify the potential cell types these cells could be differentiating into. A fluorescent microscope is used to examine and image the cells. Currently, cell counting and cell analysis are done manually. The challenges for developing accurate automated cell counting are a wide range of cells in an image given different antibodies, different cell sizes, low contrast, and cell occlusion.

The main contributions of this work are as follows.
\begin{enumerate}
    \item An annotated dataset for automated cell counting along with the domain knowledge to use the dataset. The annotation includes the cell locations as well as the count of cells per image. To the best of our knowledge, there is no annotated fluorescent microscopic cell image dataset that covers as many staining methods as this dataset.
    \item Performance comparison of the state-of-the-art regression-based and density map estimation DNN methods. The results can be used as baseline results for future improvement. The source code and the trained models are available publicly at https://github.com/ISU-NRT-D4/cell-analysis.
\end{enumerate}

The rest of the paper is organized as follows. In Section 2, we provide a summary of existing datasets related to cell counting. Section 3 presents our data collection and annotation process and the details of our new dataset. Section 4 includes applicable scenarios to utilize the dataset. Section 5 details the baseline experimental results on the dataset with five DNN models. Finally, we provide a conclusion and description of the future work in Section 6.

\section{Existing Datasets}
Several cellular image datasets are available publicly. Some datasets are for detection based methods \cite{kromp_annotated_2020, edlund_livecelllarge-scale_2021,ljosa_annotated_2012,morelli_automating_2021,parekh_evican_2022}. These datasets, consisting of either contour or smallest bounding box annotations of individual cells, are suitable for cell counting tasks. A few datasets are specifically intended for cell counting \cite{kainz_you_2015,lempitsky_learning_2010,minn_high-resolution_2020,sirinukunwattana_locality_2016}. Table \ref{table:1} summarizes the existing datasets for cell counting for different types of cells. Lempitsky and Zisserman provided a synthetic dataset of RGB images of bacterial cells from fluorescence microscopy \cite{lempitsky_learning_2010}. This dataset is widely used for training machine learning models for cell counting. Kainz et al. \cite{kainz_you_2015} introduced a dataset of brightfield microscopic bone marrow cell images. These are RGB images with inhomogeneous backgrounds. The dataset described in \cite{minn_high-resolution_2020} is comprised of histology RGB images of human embryonic cells. The images in this dataset have noisy backgrounds and a large variance in the number of cells in the images. To the best of our knowledge, the proposed dataset is the only dataset that contains images of Adult Hippocampal Progenitor Cells with different antibodies for staining.

\begin{table}
\centering
\caption{Datasets for Cell Counting}
\label{table:1}
\begin{tabular}{p{1.0in}p{1.0in}p{0.45in}p{0.45in}} 
\hline
\textbf{Dataset}               & \textbf{Type of cell}              & \textbf{No. of images} & \textbf{Resolution}  \\
\hline
\textbf{Existing}              & ~                                  & ~                      & ~                                      \\
Synthetic Bacterial Cells \cite{lempitsky_learning_2010} & Bacteria                           & 200                    & 256 x 256                              \\
Bone Marrow \cite{kainz_you_2015}                & Bone Marrow                        & 40                     & 60 x 60                                \\
Colorectal Cancer Cells \cite{sirinukunwattana_locality_2016}   & Colorectal Cancer Cells            & 100                    & 500 x 500                              \\
hESCs \cite{minn_high-resolution_2020}                     & Human Embryonic Stem Cells         & 49                     & 512 x 512                              \\ 
\hline
\textbf{Proposed}              & ~                                  & ~                      & ~                                      \\
IDCIA                          & Adult Hippocampal Progenitor Cells & 262                    & 800 x 600                              \\
\hline
\end{tabular}
\end{table}

Compared to thousands of images in the public datasets for segmentation and detection of generic objects (e.g., Microsoft COCO \cite{lin_microsoft_2015}, PASCAL VOC \cite{everingham_pascal_2010}, and CityScapes \cite{cordts_cityscapes_2016}), each of the datasets in Table \ref{table:1} has much fewer images. Moreover, there is no dataset that incorporates images from more than five antibody staining methods, as well as additional information about the antibody used per image. Having more public datasets and ground truth is desirable for automated cell image analysis.

\section{IDCIA: Proposed Image Dataset}
We describe the data collection process and annotation process and the structure of the dataset.
\subsection{Data Collection}
\begin{figure}[h]
  \centering
  \includegraphics[scale=0.5]{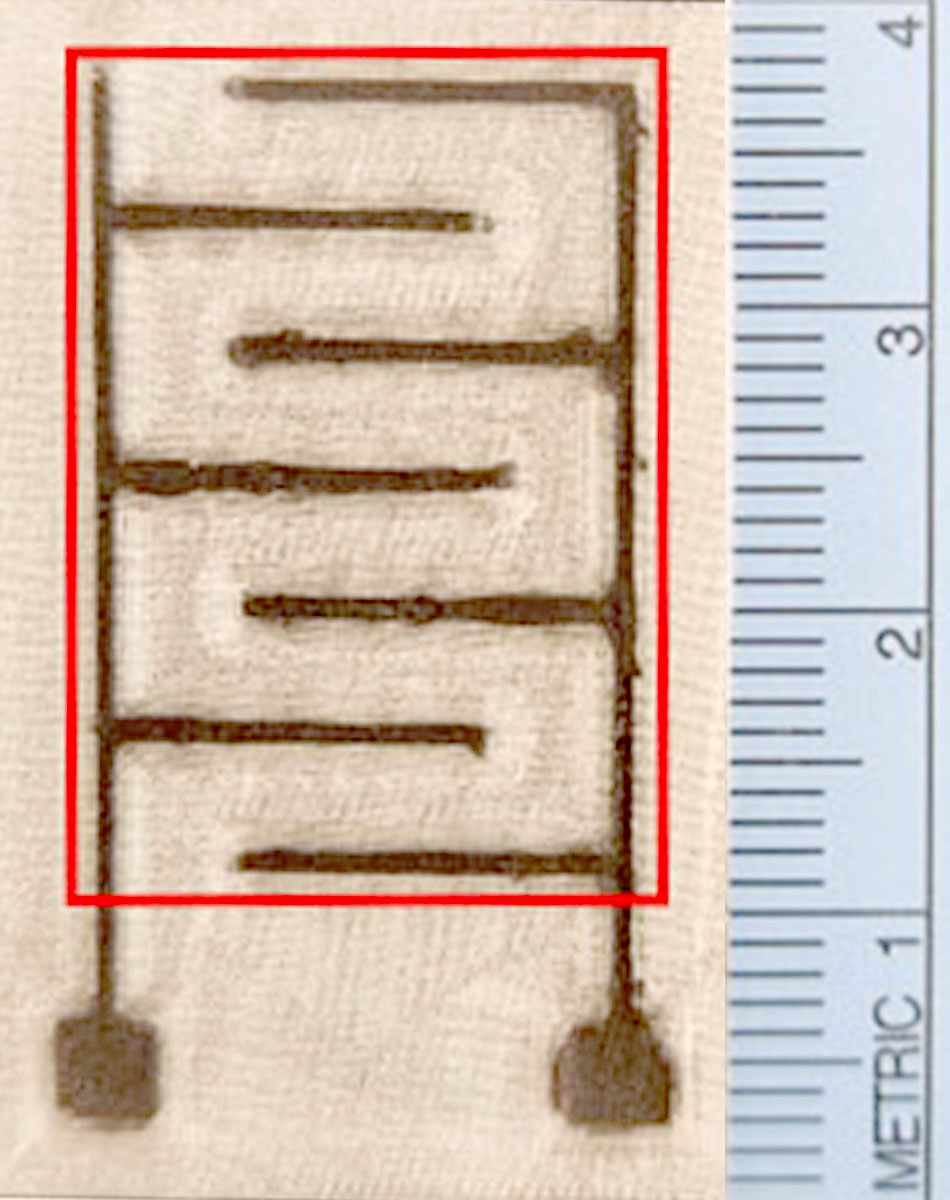}
  \caption{Scaffold for cells undergoing an electrical stimulation. The red box indicates the cell culture region.}
  \label{fig:1}
  \Description{The graphene-based scaffolds where cells are grown and electrical stimulation is applied.}
\end{figure}

Our dataset contains images of rat Adult Hippocampal Progenitor Cells (AHPCs) \cite{kirby_adult_2015} after electrical stimulation experiments and ICC. AHPCs have the potential to differentiate into the three primary cell types of the central nervous system in vitro: Neurons, Astrocytes, and Oligodendrocytes. The experiments were performed in the Sakaguchi Lab\footnote{https://faculty.sites.iastate.edu/dssakagu/} at Iowa State University. The cells were generously gifted by Dr. Fred H Gage\footnote{https://gage.salk.edu/}. The experiments started by placing 400,000 cells onto scaffolds containing graphene-based interdigitated electrode circuits \cite{uz_development_2019}. Fig. \ref{fig:1} shows a picture of 3D-printed polylactic acid (PLA) scaffolds used in the experiments. PLA is a biocompatible and biodegradable polymer making it an ideal substrate for supporting cell growth. The scaffolds were, in turn, placed inside 60mm dishes containing cell culture media. During the stimulation, the scaffolds were removed from the dish and covered with a small volume of media, and the electrode pads of the scaffolds were connected to a power supply with a desired voltage.

 For this dataset, the cultured cells underwent electrical stimulation at 125 mV for 10, 15, or 20-minute durations, once a day for a period of 7 days. An additional scaffold was set aside as a control and received no stimulation. After the 7-day period, ICC was performed on both the stimulated and non-stimulated samples to evaluate various neural differentiation markers. The process occurred as follows. The electrode pad regions of the scaffolds were discarded, and then each scaffold was cut into six pieces using sterile scissors. The cut pieces then underwent a process of ICC that involved a fixation process, repeated rinsing, and incubation in primary and secondary antibodies. Primary antibodies bind to antigens, whereas fluorophore-conjugated secondary antibodies bind to a primary antibody to allow for the indirect detection of the target protein. Cell nuclei were stained with DAPI. Table \ref{table:2} shows the seven primary antibodies used during the experiments. Once this process was complete, the pieces were mounted onto microscope slides for fluorescence imaging with an upright fluorescence microscope (Nikon Microphot FXA) for visualization. Imaging of the cells was made with a 20x objective, and images were captured with a CCD camera. Fig. 2(A-C) shows pseudo-colorized examples of the cells.

Imaging and counting of cells of  the microscope images were conducted blind. That is, the individuals conducting these processes were not informed which scaffolds underwent which stimulation condition. This was done to limit the amount of bias when collecting the results of the experiments.
\begin{table}
\centering
\caption{Seven primary antibodies used in the experiments}
  \label{table:2}
\begin{tabular}{ll} 
\hline
\textbf{Antibody} & \textbf{Cell Type Identification}  \\ 
\hline
DAPI     & Cell Nuclei                        \\
TuJ1              & Immature Neurons                   \\
MAP2ab            & Maturing Neurons                   \\
RIP               & Oligodendrocytes                   \\
GFAP              & Astrocytes                         \\
Nestin   & Neural stem cells                  \\
Ki67              & Proliferating Cells                \\
\hline
\end{tabular}
\end{table}

 Following the imaging of the samples, manual annotation was performed by a group of undergraduate students led by a graduate student with more than three years of experience in cellular image analysis. The ImageJ \cite{schneider_nih_2012} Cell Counter tool was used to annotate a cell by manually placing a dot on each cell in an input image. The tool reports the total number of dots in the image. ImageJ is a powerful tool for processing and analyzing scientific images. The tool also allows the measurement of various cell properties, such as size, shape, and intensity.

 We used dot annotations to label the cells for counting purposes. First, dot annotation enables the accurate marking of individual cells. This is to avoid double counting some cells or missing to count other cells for images with a large number of cells or with densely packed or overlapping cells. Second, dot annotation is a fast and efficient way to count and identify cells since the exact cell contour or bounding box is not required. It is useful for expanding the dataset in the future. Third, dot annotations can be used to verify how a DNN model arrives at the predicted cell count, which should improve cell biologists’ trust in the model. Finally, dot annotations are useful for the development and evaluation of both detection-based and regression-based DNN methods for cell counting. 

 \subsection{Dataset Structure and Details}

After the completion of annotating all the images, we split the dataset into three non-overlapping sets: Training, Validation, and Testing at the ratio of 60:20:20. To ensure that each of these sets contains a proportional number of samples from each antibody type, we used stratified sampling based on antibody type. We then ran a program to extract the coordinates of individual dots from a dot-annotated image by thresholding the color of the dot and saving the coordinates in a csv file.

Table \ref{table:3} presents statistics about the IDCIA dataset of 262 images with 84 cells on average per image. Notice that the cells are not evenly distributed across antibodies. Images from DAPI staining to identify cell nuclei has the most cells. The dataset has a high variance in terms of the number of cells per image. This dataset brings an interesting and challenging problem due to the variability introduced by the use of multiple antibodies in immunolabeling experiments. Each antibody interacts differently with the cells, resulting in a wide range of appearances and visual characteristics within the same sample. This can make it difficult to accurately and reliably count the number of cells present. Additionally, the presence of multiple antibody labels may also result in overlap between cells, further complicating the cell counting process.

\begin{table}
\centering
\caption{Statistics of IDCIA: 800x600 image resolution; range of the number of cells per image [0, 581]}
  \label{table:3}
\begin{tabular}{llp{1.5in}l} 
\hline
\textbf{Antibody} & \textbf{No. of images} & \textbf{Mean cell count per image ± std}  \\ 
\hline
DAPI              & 119                    & 141.35 ± 122.56                           \\
TuJ1              & 25                     & 17.48 ± 18.34                             \\
MAP2ab            & 24                     & 49.71 ± 35.07                             \\
RIP               & 24                     & 49.542 ± 45.65                            \\
GFAP              & 23                     & 3.43 ± 5.17                               \\
Nestin            & 23                     & 89.35 ± 79.70                             \\
Ki67              & 24                     & 8.17 ± 8.38                               \\
\textbf{Total}    & \textbf{262}           & \textbf{83.857±104.42}                    \\
\hline
\end{tabular}
\end{table}

\begin{figure*}
  \includegraphics[width=\textwidth ,scale=0.9]{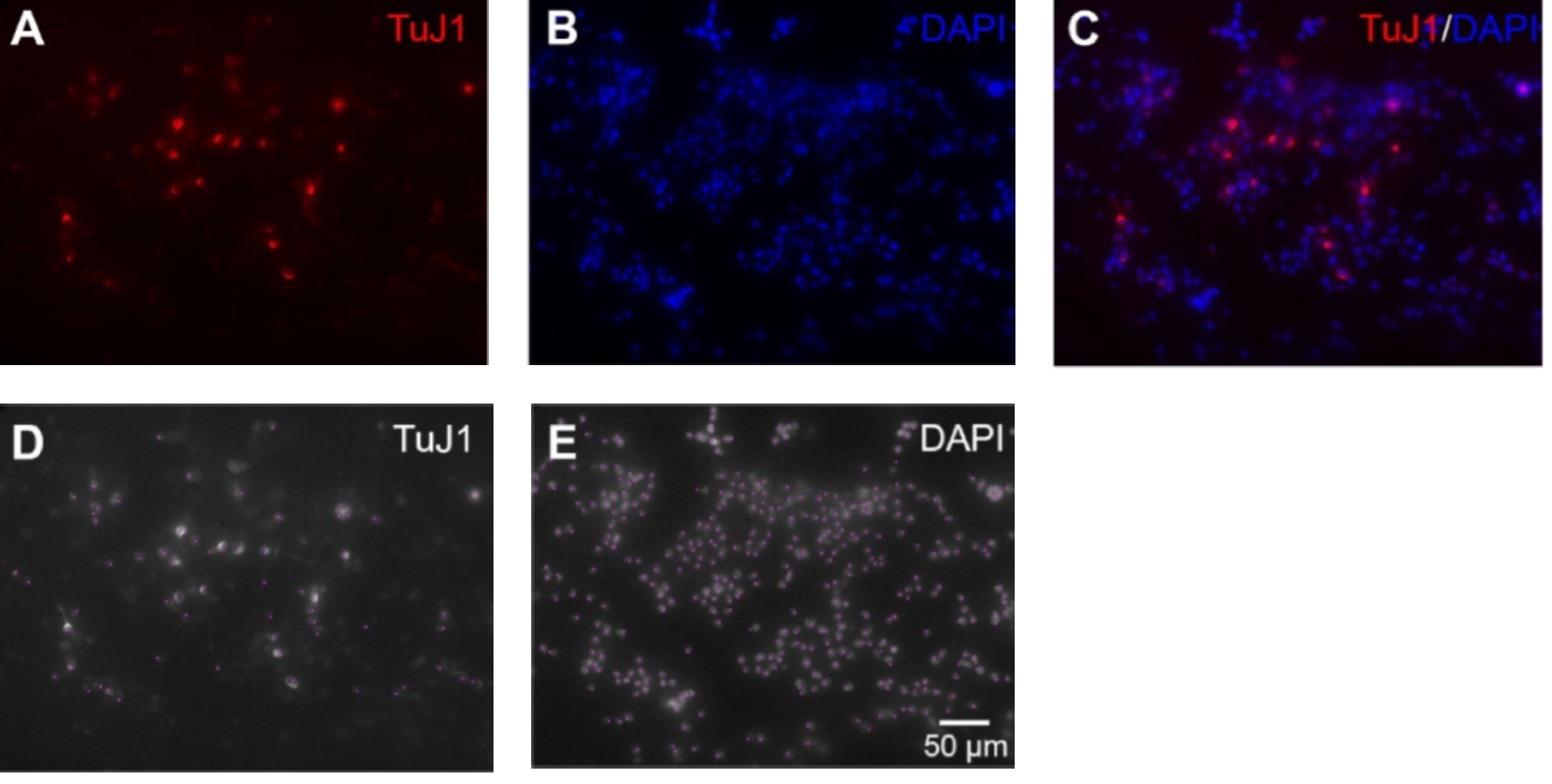}
  \caption{Quantification of immature neurons in AHPCs after 7 days in vitro (DIV) of electrical stimulation. Row 1 (A-C) shows fluorescence images of AHPCs labeled with an immature neuron marker (TuJ1, red; A and C) and nuclei marker (DAPI, blue; B and C) following 15 min. of 125 mV electrical stimulation once a day for 7 days. Row 2 (D-E) shows the dot-annotated images of the TuJ1 (D) and DAPI (E) staining by using the ImageJ Cell Counter tool to put a pink dot on a cell to be counted. Scale bar = 50 µm. Images have been pseudo-colorized for better visualization.}
  \label{fig:teaser}
\end{figure*}

The provided dataset has two directories: \textit{images} and \textit{ground\_truth}, and the \textit{readme.md} file for the description of the dataset folder. The images directory contains seven sub-directories, one for each of the primary antibodies used in the experiments. Inside each directory are grayscale images from fluorescence microscopy. The images were resized to 800x600 pixels. The images in Fig. \ref{fig:teaser} were pseudo-colored for better visualization. All the images are named using a consistent format that gives information about the antibody, objective, cell type, and field number. Similarly, the ground truth directory contains seven directories under it. Inside these directories are .csv files containing coordinates of the dot-annotated cells for each image in the images directory. For the reproducibility of experiments, we provide .csv files that list the names of the images in the training, validation, and testing sets.

\section{Usage Scenarios of IDCIA}
The dot-annotated microscopic cell images and accompanying immunolabeling and staining information provided in this dataset offer opportunities for computer scientists to contribute to advancing cell biology research. Here, we outline a few potential applications.

The dot annotations with the number and location of cells marked for each image are useful for developing effective and interpretable counting methods, as outlined in Section 2. As cell counting is an important task in cell analysis, desirable automated methods should produce the predicted count within the experts’ acceptable error rate of, at most 5\% difference from the actual cell count. If automated cell counting is quick, accurate, and trustworthy, electrical stimulation experiments for stem cell therapy can be accelerated. Due to a high variance in the number of cells labeled for each staining antibody, the name of the antibody used may be useful for improving automated cell counting methods. For instance, the DAPI staining of cell nuclei identifies many more cells than the immunolabeling with the antibodies. This dataset includes images from the experiments covering the use of seven primary antibodies for immunolabeling. On the contrary, DNN methods may be developed to classify images to predict the antibody used for cell labeling. We refer to this problem as antibody classification problem.

Given the limited number of datasets for cellular image analysis and the limited number of images for each dataset, more high-quality datasets are needed. Our dataset supplements existing datasets and may be useful for transfer learning that extracts information from data in one domain and transfers the learned knowledge to another domain \cite{lavitt_deep_2021,shao_transfer_2015}.

Both the cell count and the locations of individual cells are useful for developing cell segmentation or cell detection methods based on high-level labeling (i.e., weak supervision). Dot annotations, which are faster to acquire than detailed annotations, can be leveraged to obtain more detailed annotations and reduce manual labeling time. Utilizing ground truth information obtained through weak supervision can help in the automated segmentation and detection of individual cells. This allows the measurement of cell shape and orientation. Currently, proprietary software such as MetaXpress exists for measuring cell shape and cell orientation. However, the software requires that cell culture be done on a smooth surface, which greatly limits the opportunities for using custom-designed scaffolds for electrical stimulations, as shown in Fig. \ref{fig:1}.

\subsection{Suggested Metrics}
Suggested metrics for the cell counting task are Mean Absolute Error (MAE) \cite{bishop_pattern_2006} and Root Mean Squared Error (RMSE) \cite{bishop_pattern_2006}. MAE is the average of the absolute difference between the label ground truth count $y_i$ and the predicted value count $\hat{y_i}$ for all n images in a given dataset. RMSE penalizes large errors to a greater extent compared to MAE. See Equations 1-2. We   introduce Acceptable Error Count Percent (ACP) to measure the percentage of images whose predicted count is within a 5\% difference from the true count by the domain expert, as shown in Equation 3. We use Iverson brackets $\llbracket . \rrbracket$ to denote a function that returns 1 if the condition is satisfied or 0 otherwise. These metrics are calculated below. Due to limited space, we only report MAE and ACP in this paper.

\begin{equation}
MAE = \left(\frac{{1}}{{n}}\right)\sum_{{i=1}}^{{n}}\left|{\hat{y}}_{i}{-} y_i\right|
\end{equation}

\begin{equation}
RMSE=\sqrt{\left(\frac{1}{n}\right)\sum_{i=1}^{n}\left({\hat{y}}_\mathrm{i}-y_i\right)^2}
\end{equation}

\begin{equation}
ACP=\left(\frac{1}{n}\right) \ast 100 \sum_{i=1}^{n} \llbracket |\hat{y_i} - y_i | \leq 0.05 \ast y_i \rrbracket
\end{equation}
The lower the values of MAE, the more accurate a model's predictions are. On the other hand, a higher ACP indicates that more predictions are in the acceptable margin. For the antibody classification problem, traditional performance metrics for classification problems such as accuracy, precision, recall, and F1-score can be used \cite{bishop_pattern_2006}.

\section{Baseline Experiments for Cell Counting}

We evaluated five different models for cell counting using the  IDCIA dataset. They are a Convolutional Neural Network (CNN) with regression output (CNN Regression), two-crowd counting methods (CSRNet \cite{li_csrnet_2018}, MCNN \cite{zhang_single-image_2016}), and two cell-counting methods (Count-ception \cite{paul_cohen_count-ception_2017}, FCRN \cite{xie_microscopy_2018}). All these four methods are based on density map estimation. For CSRNet and Count-ception, we used the source code provided by the original authors of the methods. Our CNN Regression model has a pre-trained VGG16 \cite{simonyan_very_2015} network with a fully connected layer at the end of it, followed by one output neuron for the predicted cell count for a given image. We excluded \cite{morelli_automating_2021,khan_deep_2016} from our experiments since they require detailed annotations for training. All models were implemented in Python using the PyTorch \cite{paszke_pytorch_2019} library and trained on NVIDIA Tesla T4 and P2000 GPUs.

Count-ception and FCRN were developed for cell counting. Count-ception is a network of fully convolutional layers without any pooling layer. This is to avoid losing pixel information and to ease the calculation of the receptive field. Given an input image, Count-ception produces an intermediate count map. Each network inside it counts the number of objects in its receptive field. FCRN uses CNN to regress a cell’s spatial density across an image. It first maps the input image to feature maps with dense representation and then recovers the spatial span by bilinear up-sampling. FCRN allows prediction for input with an arbitrary size. FCRN-A is a version of FCRN that uses small 3 × 3 kernels for every convolutional layer, and each convolutional layer is followed by a pooling layer.

CSRNet \cite{li_csrnet_2018} and MCNN \cite{zhang_single-image_2016} are density-map based models developed for counting people in a congested environment. These models can handle dense crowds, which makes them well-suited for handling cell congestion. In addition, these models were designed to be robust to variations in object size and shape, lighting, and contrast conditions. CSRNet is a two-component network with a CNN as the first component for feature extraction. The second component is a dilated CNN to produce larger reception fields, replacing pooling operations. MCNN extracts scale-relevant features by using filters with different sizes of receptive fields. The authors proposed a network of three parallel CNNs with different filter sizes. For an input image, the network averages the predicted density maps of the three CNNs and outputs a final count prediction. To use the dot-annotated images for training CSRNet and MCNN, we followed the ground truth generation method in \cite{zhang_single-image_2016} by blurring each dot annotation using a Gaussian kernel to produce corresponding density maps. Since the generated density maps have a high impact on the performance of the models for cell counting, we used geometry-adaptive kernels \cite{li_csrnet_2018} to accurately generate corresponding density maps for input images. Fig. 3 (left) shows input images. Fig. 3 (right) shows the corresponding density maps generated.

\begin{figure}[h]
  \centering
  \includegraphics[scale=0.65]{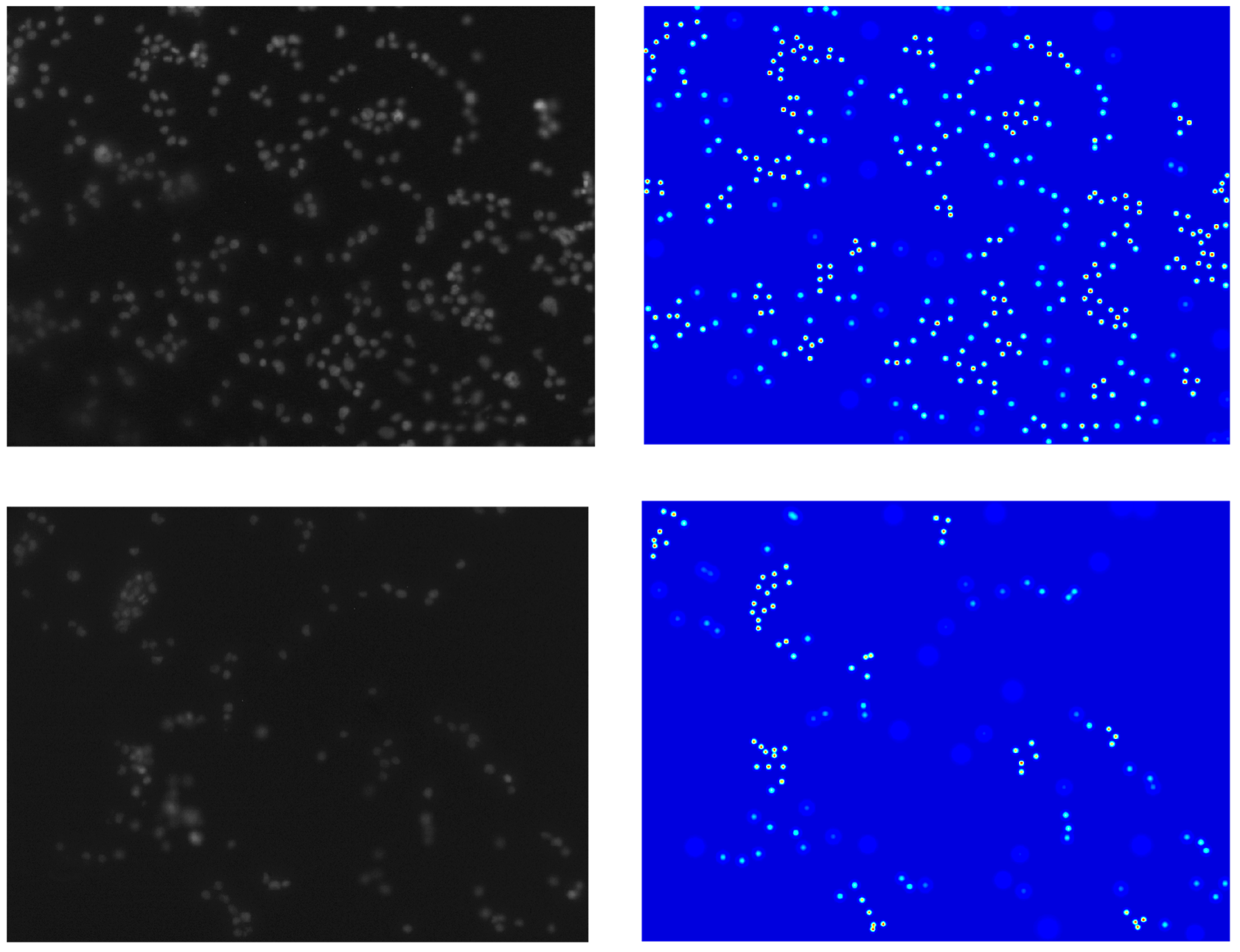}
  \caption{Images (left) and generated density maps (right)}
  \label{fig:3}
  \Description{images on the left and Density maps on the right}
\end{figure}

All models were then trained using an end-to-end stochastic gradient descent method \cite{ruder_overview_2017} and data augmentations per the original authors’ code. The loss function used in training was PyTorch L1Loss. The grid search method was conducted to obtain the best hyperparameter values on the validation dataset. For each method, we performed five runs on the IDCIA dataset.  Each run involved training the model using the hyperparameter values that give the best MAE on the validation dataset. The average of the MAEs over the five runs was reported for each model.

\begin{table}
\centering
\caption{Avg. MAEs of the five methods on the IDCIA test dataset}
\label{table:4}
\begin{tabular}{p{1.0in}lp{0.2in}lp{0.2in}lp{0.2in}lp{0.2in}l} 
\hline
\textbf{Model}              & \textbf{Learning rate} & \textbf{Batch Size} & \textbf{Epochs} & \textbf{Avg. MAE}  \\ 
\hline
CNN Regression              & 3.5e-05                   & 16                  & 500             & 24.15              \\
CSRNet \textbf{\cite{li_csrnet_2018}}        & 2e-7                   & 1                   & 400             & 18.64              \\
MCNN \textbf{\cite{zhang_single-image_2016}}          & 1e-7                   & 1                   & 1500            & 19.57              \\
Count-ception \textbf{\cite{cohen_count-ception_2017}} & 1e-2                   & 2                   & 1000            & \textbf{15.47}     \\
FCRN-A \textbf{\cite{xie_microscopy_2018}}        & 1e-2                   & 16                  & 1000            & 27.49              \\
\hline
\end{tabular}
\end{table}

The optimal learning rate, batch size, and the number of epochs, as determined by the grid search method, are given for each counting method. The results are indicated in Table \ref{table:4}. Count-ception is the best method giving the lowest MAE. CSRNet and MCNN perform comparably despite being originally proposed for crowd-counting tasks. CNN Regression and FCRN-A are the two worst methods. FCRN-A gives the worst performance on IDCIA, although it was proposed for cell counting. FCRN-A uses small 3 × 3 kernels for every convolutional layer, and each convolutional layer is followed by a pooling layer.

According to Table \ref{table:5}, the CSRNet model has the highest ACP at 17\%, which indicates that it has the best performance. On the other hand, the MCNN model has the lowest ACP at 0\%, indicating poor performance in this aspect, even though it has comparable performance under the mean absolute error (MAE) metric. The CNN regression, Count-ception, and FCRN-A models have the same ACP at 9\%, indicating similar performance. While MCNN performs decently in terms of MAE, it performs the worst based on ACP, which is an important criterion for domain experts.

\begin{table}
\centering
\caption{Comparison of ACP of the five methods}
\label{table:5}
\begin{tabular}{ll} 
\hline
\textbf{Model}           & \textbf{ACP}\%  \\ 
\hline
CNN Regression & 9\%    \\
CSRNet          & 17\%   \\
MCNN            & 0\%    \\
Countception    & 9\%    \\
FCRN\_A         & 9\%    \\
\hline
\end{tabular}
\end{table}

Breaking down the performance of the models with respect to the different antibody labeling in Table \ref{table:6}, Count-ception performs better on most of the staining types, while the simple CNN Regression model exhibits the best performance in two of the staining methods. CSRNet performs best for the DAPI labeled samples, while MCNN performs best for the Ki67 labeled samples. Such performance differences are caused by the different visual appearances of cells under different staining antibodies. It indicates the need to build models that can produce an accurate prediction for a given image under a certain antibody labeling. Count-ception has the lowest MAE on the TuJ1, MAP2ab, and Nestin immunolabeled images. Count-ception is the second best for the rest of the antibody labels. CNN Regression has the lowest MAE on the RIP and GFAP labeled images. 

Overall, our results demonstrate the potential of DNNs for counting cells from microscope images. IDCIA has a high variance in terms of cell count for different antibody labels. In our experiments, Count-ception undercounts cells in some images but overcounts in others. It undercounts all the DAPI labeled images in the test dataset. This highlights the importance of evaluating the performance of different models on different immunolabeling antibodies.

\begin{table}[]
\centering
\caption{Avg. MAE per antibody labeling on the IDCIA test dataset}
\label{table:6}
\begin{tabular}{p{0.5in}lp{0.3in}lp{0.3in}lp{0.3in}lp{0.3in}lp{0.3in}l}
\hline
\textbf{Staining} & \multicolumn{5}{l}{Avg. MAE by   different methods}      \\ \cline{2-6} 
                          & \textbf{CNN Reg.} & \textbf{CSRNet } & \textbf{ MCNN}  & \textbf{Count-ception }& \textbf{ FCRN\_A} \\ \hline
DAPI                      & 25.94    & \textbf{13.29}  & 29.84 & 15.8          & 21.82   \\
TuJ1                      & 16.98    & 7      & 4.66  & \textbf{3.11}          & 11.23   \\
MAP2ab                    & 37.72    & 41.4   & 46.59 & \textbf{25.2}          & 38.13   \\
RIP                       & \textbf{14.43}    & 15.2   & 23.42 & 16.9          & 64.43   \\
GFAP                      & \textbf{36.13}    & 75.4   & 37.15 & 38.59         & 58.67   \\
Nestin                    & 19.05    & 15.5   & 9.08  & \textbf{6.69}          & 8.41    \\
Ki67                      & 16.32    & 7.4    & \textbf{1.11}  & 2.18          & 7.56    \\ \hline
\end{tabular}%
\end{table}

\section{Conclusion and Future work}
In this paper, we present a new annotated dataset of images of cells from a fluorescence microscope. The cells were immunolabeled using a panel of cell type-specific antibody markers, and all cell nuclei stained using DAPI. The dataset is available for public use along with the source code and the trained models. We present the effectiveness of deep-learning methods for counting on the dataset. We found that different existing deep-learning models are best for different antibodies used for labeling. All the methods still underperform when using the ACP metric based on the domain experts, leaving room for improvement. The results of our study highlight the challenges in accurately predicting cell counts. We plan to continue to explore different architectures and training techniques in order to increase the performance on the ACP metric. Future work includes the development of a new DNN method that is sufficiently accurate and acceptable by domain experts.

\begin{acks}
    We thank all the undergraduate students, John Swanson, Gabrielle Sawin, and Anna Garbe, who participated in the imaging and annotation of the fluorescence microscopy images. This work is partially supported by the NSF Grant No. 2152117. Findings, opinions, and conclusions expressed in this paper do not necessarily reflect the view of the funding agency.
\end{acks}

\bibliographystyle{ACM-Reference-Format}
\bibliography{refs}

\end{document}